\documentclass[%
 reprint,
 amsmath,amssymb,
 aps,
prl,
superscriptaddress,
longbibliography
]{revtex4-1}

\usepackage{graphicx}
\usepackage{dcolumn}
\usepackage{bm}

\newcommand{\beginsupplement}{%
        \setcounter{table}{0}
        \renewcommand{\thetable}{S\arabic{table}}%
        \setcounter{figure}{0}
        \renewcommand{\thefigure}{S\arabic{figure}}%
     }

\pdfoutput=1

\begin{document}
\title{Experimental realization of optimal energy storage in resonators embedded in scattering media}


\author{Philipp del Hougne}
\affiliation{Institut de Physique de Nice, CNRS UMR 7010, Université C\^ote d’Azur, 06108 Nice, France}
\author{Remi Sobry} 
\affiliation{Univ Rennes, CNRS, Institut d’Electronique et de Telecommunications de Rennes, UMR–6164, F-35000 Rennes, France}
\author{Olivier Legrand}
\affiliation{Institut de Physique de Nice, CNRS UMR 7010, Université C\^ote d’Azur, 06108 Nice, France}
\author{Fabrice Mortessagne}
\affiliation{Institut de Physique de Nice, CNRS UMR 7010, Université C\^ote d’Azur, 06108 Nice, France}
\author{Ulrich Kuhl}
\affiliation{Institut de Physique de Nice, CNRS UMR 7010, Université C\^ote d’Azur, 06108 Nice, France}
\author{Matthieu Davy}
\affiliation{Univ Rennes, CNRS, Institut d’Electronique et de Telecommunications de Rennes, UMR–6164, F-35000 Rennes, France}

\date{\today} 

\begin{abstract}
The ability to enhance light-matter interactions by increasing the energy stored in optical resonators is inherently dependent on their coupling to the incident wavefront. In practice, weak coupling may result from resonators' irregular shapes and/or the scrambling of waves in the surrounding scattering environment. Here, we present a non-invasive wavefront shaping technique providing optimal coupling to resonators. The coherent control of the incident wavefront relies on the lengthening of delay times of waves efficiently exciting the resonator. We demonstrate our concept in microwave experiments by injecting in-situ optimal wavefronts that maximize the energy stored in multiple high-permittivity dielectric scatterers and extended leaky cavities embedded in a complex environment. We expect our framework to find important applications in the enhancement of light-matter interactions in photonic materials as well as to enhance energy harvesting.
\end{abstract}


\maketitle

Light incident upon a photonic resonator can be efficiently trapped by a long-lived mode if the light's frequency is within a narrow interval around the resonance. Since the energy stored in the resonator is proportional to the light's dwell time \cite{Smith1960,Lagendijk1996,rotter2017light,Durand2019}, light confinement in photonic resonators constitutes an important mechanism to enhance light-matter interactions, for instance, to generate non-linear optical effects. 
Photonic resonators can take the form of optical microcavities \cite{Chang1996,Vahala2003,Liu2013,Cao2015}, nanocavities \cite{Akahane2003,Song2005} or Anderson-localized modes in disordered crystals \cite{Sapienza2010}, to name a few examples, and are also crucial to boost the absorption rate in light harvesting schemes \cite{Yablonovitch1982b,Vynck2012,Yao2012,Garin2014}.
However, coupling light incident from the far-field to an optical resonator is a major challenge in many practical scenarios where (i) the resonator's shape is unknown or irregular and/or (ii) the resonator is embedded in a complex scattering environment. The latter completely scrambles the incident wavefront such that its coupling to the resonator, and consequently the energy storage, is dramatically reduced.

To counteract the effects of this scrambling, many wavefront shaping (WFS) techniques have been developed within the last decade that rely on tailoring the wavefront incident on a complex medium to coherently control wave propagation within the medium \cite{Mosk2012,rotter2017light}. In its simplest form, WFS may enhance energy storage in a point-like resonator embedded in a complex medium by focusing the wave field on its location. To determine how the incident wavefront should be shaped, such schemes must access in some way information about the wave field at the resonator's location. To circumvent the need for direct field measurements, a number of proposals indirectly obtain this information by implanting a guide-star at the target location \cite{Vellekoop2008,Horstmeyer2015}, by creating a virtual guide-star with multi-wave approaches \cite{Larrat2010,Chaigne2014}, by relying on a non-linear response at the target position \cite{Katz2014,Hougne2017} or on parametric variation of the target \cite{Ma2014,Zhou2014,Ambichl2017b,Horodynski2019}. 

An alternative approach to couple energy into an embedded resonator without relying on any of the above-described conditions, and moreover also applicable to extended resonators, is related to the impact of the resonator's presence on the dwell time of waves that interacted with it. The Wigner-Smith time-delay operator (WSO) provides a tool to extract an incoming wavefront optimizing the delay time \cite{Rotter2011,Carpenter2015,Gerardin2016,Xiong2016,Ambichl2017a,Bohm2018}. As long as the resonator's quality factor is clearly superior to that of the surrounding medium, the eigenstate of the WSO associated with the largest delay time may strongly increase the energy stored in the resonator \cite{Durand2019}. In this article, we provide an experimental demonstration of this concept in the microwave domain. By injecting the optimal wavefronts in-situ, we observe a corresponding enhancement of the stored energy for single or multiple dielectric cylinders as well as for an extended leaky cavity, each embedded in a complex scattering environment. We discuss the theory behind the optimality of the approach and highlight its limitations when the quality factors of resonator and medium become comparable.

\section*{Theory}

\subsection*{Wigner-Smith time-delay operator} The delay time of waves travelling through the medium carries key information about non-cooperative resonant inclusions. By identifying the wavefront that maximizes the delay time between incoming and outgoing waves, these time-delay signatures of embedded resonators can be leveraged to optimally excite the resonators and store energy within their volume. The delay time of outgoing waves $E_o=S(\omega)E_i$ for an incoming wavefront $E_i$ is formally given by \cite{Durand2019}

\begin{equation}\label{Eq_time_wavefront}
\tau(\omega,E_i)=-i\frac{E_o^\dagger \frac{\partial E_o}{\partial\omega}}{||{E_o}||^2}=-i\frac{E_i^\dagger S^\dagger(\omega) \frac{\partial S(\omega)}{\partial \omega}E_i}{E_i^\dagger S^\dagger(\omega) S(\omega) E_i}.
\end{equation}

\noindent The scattering matrix $S(\omega)$ gives the fullest account of transmitted and reflected field coefficients between the channels coupled to the system. $S(\omega)$ can be decomposed into transmission (TM) and reflection (RM) matrices as 
\begin{equation} 
S= 
\begin{pmatrix} 
r & t^T\\
t & r'
\end{pmatrix}.
\quad
\label{SM}
\end{equation}
For systems with flux-conservation, the scattering matrix is unitary, $S^\dagger S=I$, so that for a normalized incoming wavefront the definition in Eq.~\ref{Eq_time_wavefront} coincides with the delay time $\tau(\omega,E_i)=E_i^\dagger Q(\omega) E_i$ found using the WSO \cite{Wigner1955,Smith1960,Brouwer1997}

\begin{equation}
Q(\omega)=-iS^{-1}(\omega) \frac{\partial S(\omega)}{\partial \omega}.
\end{equation}
The eigenvalues of $Q(\omega)$ verifying $Q(\omega)q_n=\tau_n {q}_n$ are known as the proper delay times. They are real and give well-defined delay times obtained upon using the eigenvectors of $Q(\omega)$ as incident wavefronts: $\tau_n=\tau(\omega,q_n)$. The eigenvector of the WSO associated with the largest proper delay time hence provides the incoming wavefront optimizing the delay time and the optimal energy stored within the medium \cite{Durand2019}. However, in many experimental setups, the energy can only be injected within the medium from one side. The WSO is then constructed from a measurement of the TM or RM. $t(\omega)$ and $r(\omega)$ are non-unitary matrices so that the corresponding WSOs,  $Q_t(\omega)=-it^{-1}(\omega) \frac{\partial t(\omega)}{\partial \omega}$ and $Q_r(\omega)=-ir^{-1}(\omega) \frac{\partial r(\omega)}{\partial \omega}$ respectively, are non-Hermitian with complex eigenvalues $\tilde\tau_n$. Neverthless, the real part of $\tilde\tau_n$ gives the frequency derivative of a scattering phase related to a delay time \cite{fan2005principal,Bohm2018}. The imaginary part of $\tilde\tau_n$ reflects the variation of transmitted or reflected intensities with frequency.

\subsection*{Optimality} Let us take the example of $Q_r(\omega)$ to demonstrate that the WSO eigenstate with the largest eigenvalue is the optimal wavefront for maximal coupling to an embedded resonator. We approach this proof from a modal perspective. The quasi-normal modes (referred to as modes in the following) of the system are the eigenfunctions $\phi_m(\boldsymbol{r})$ that are solutions of the wave equation with outgoing boundary conditions, $\Delta \phi_m(\boldsymbol{r})+\epsilon(\boldsymbol{r})(\tilde\omega_m/c_0^2)\phi_m(\boldsymbol{r})=0$, where $\epsilon(\boldsymbol{r})$ is the spatial distribution of the permittivity and $c_0$ is the speed of light. The eigenfunctions are associated with spectral resonances characterized by complex eigenfrequencies $\tilde\omega_m=\omega_m-i\Gamma_m/2$. $\omega_m$ is the central frequency and the linewidth $\Gamma_m$ is inversely proportional to the modal decay rate $\tau_m=2/\Gamma_m$ or equivalently to the quality factor $Q_m=2\omega_m/\Gamma_m$. In our case, the modes can be separated into two categories: (i) short-lived modes (small quality factors) of the surrounding environment with eigenfunctions which extend throughout the system and weakly interact with the resonator; and (ii) long-lived modes spatially localized on the resonator whose quality factors significantly exceed those of the first category. 

We analyze the time-delay eigenstates of $Q_r(\omega)$ near the resonance with the $n$th long-lived resonator mode, $\omega\sim\omega_n$. The RM $r(\omega)$ can be decomposed as the superposition of a non-resonant contribution associated with short-lived modes, $r_0(\omega)$, and a resonant modal term with Lorentzian line expressed as $r_n(\omega)=-iW_n W_n^T(\omega-\tilde\omega_n)^{-1}$ \cite{Davy2019}, such that 

\begin{equation}
r(\omega)=r_0(\omega)-i\frac{W_n W_n^T}{\omega-\tilde\omega_n}.
\end{equation}

\noindent The vector $W_n$ of dimension $N\times1$ is the projection of the corresponding eigenfunction $\phi_n(\boldsymbol{r})$ onto the channels. The optimal wavefront $E_{\mathrm {opt}}$ is the complex conjugate (or time-reversed version) of the modal coupling vector $W_n$: $E_\mathrm{opt}=W_n^*/||W_n||$. This wavefront indeed provides maximal excitation of the strength of this mode \cite{Davy2019}. An accurate decomposition of the RM's spectra into modal contributions would directly provide the matrix $r_n(\omega)$ and hence $E_\mathrm{opt}$. However, for open systems as in the present work, the large degree of modal overlap resulting from the presence of short-lived modes precludes such a modal analysis. Such analysis is also not possible for quasi-monochromatic measurements. 

Assuming that the frequency variation of the non-resonant term is small, $r_0(\omega)=r_0$, we demonstrate in the SI appendix that $Q_r(\omega)$ is of rank one so that the left eigenvector veryfing $q_n^TQ_r(\omega)=\tilde\tau_n q_n^T$ is given by $q_n=W_n/||W_n||$. Its conjugate $q_n^*$ can therefore be identified as the optimal coupling vector between incoming channels and the resonator. The associated eigenvalue at the resonance $\omega=\omega_n$ is then equal to 

\begin{equation}
\tilde{\tau}_n=\frac{2}{\Gamma_n} \frac{\kappa(\omega_n)}{1+\kappa(\omega_n)},
\end{equation}

\noindent where $\kappa(\omega_n)=\mathrm{Tr}(r_0^{-1}r_n(\omega))=-2W_n^Tr_0^{-1}W_n/\Gamma_n$. When the non-resonant contribution is small with $|\kappa| \gg 1$, the real part of the largest eigenvalue can be identified as the delay time associated with the mode, $\tilde\tau_n=2/\Gamma_n$. For lower $|\kappa|$, the physical interpretation of the complex eigenvalue is not straightforward but the eigenvector still ensures optimal coupling to the resonator. 

\section*{Experimental Demonstration}

\begin{figure*}
\centering
\includegraphics[width=14cm]{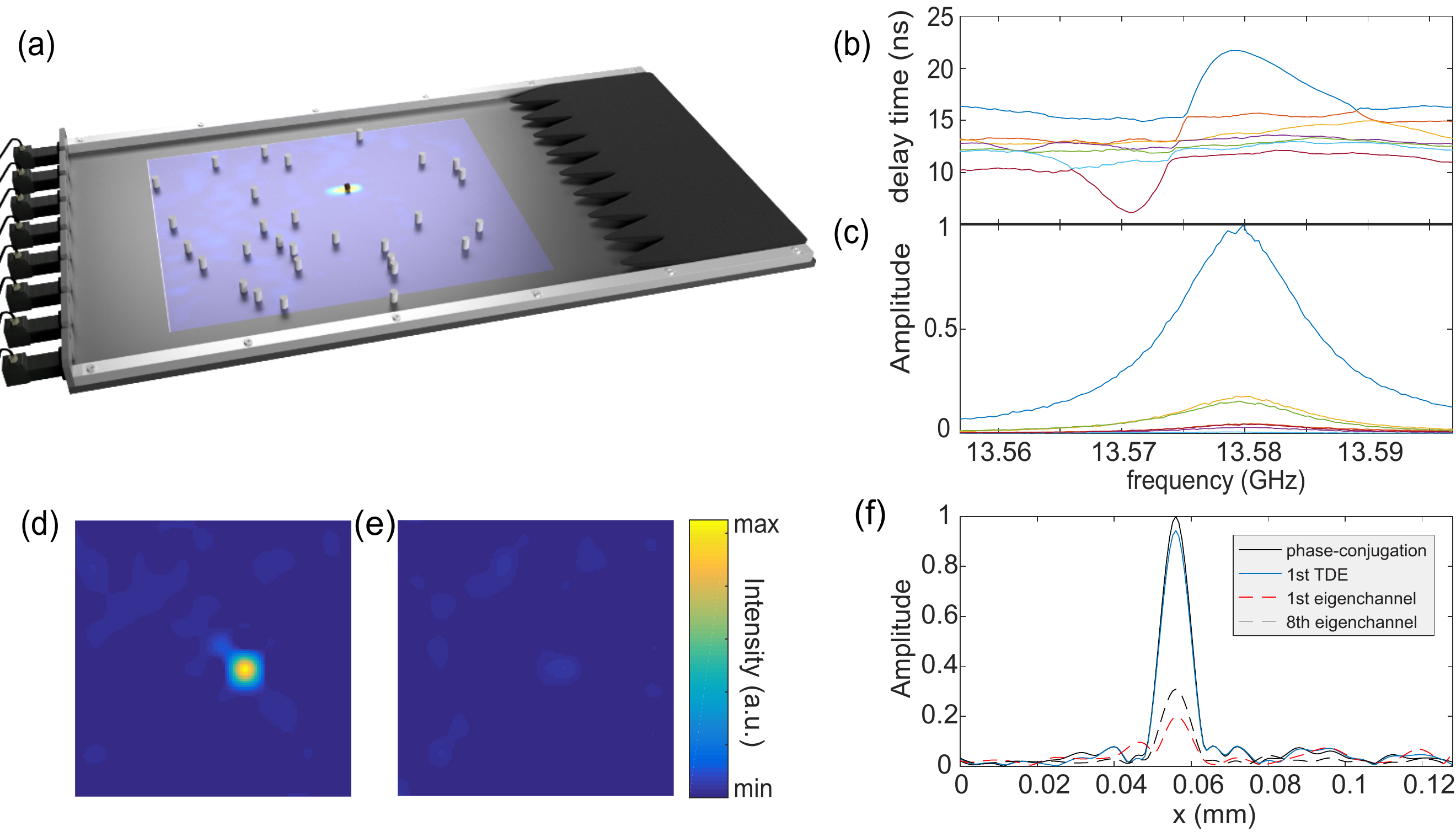}{\centering}
\caption{\label{fig1}(a) Schematic of the experimental setup. The quasi-2D waveguide's top plate has been removed to show its interior. 
$N=8$ coax antennas capable of simultaneously injecting and receiving waves are positioned on the waveguide's left end, while its right end is filled with absorbing foam. 
Randomly placed teflon scatterers (diameter 5 mm, refractive index 2.1) inside the waveguide surround a resonant dielectric disk. The projected intensity map shows the blind focusing on this embedded resonator achieved upon injecting the first time-delay eigenstate of the reflection matrix.
(b) Real part of the time-delay eigenvalues $\tilde{\tau}_n(\omega)$ associated with the eigenstates of the WSO. A peak on the first delay time is observed at $f_0=13.58$ GHz. 
(c) Spectra of the time-delay eigenstates at the dielectric resonator's location measured upon injecting in-situ the eigenvectors of the WSO computed at $f_0$. A strong enhancement corresponding to the first time-delay eigenstate is observed.
(d) Spatial distribution of energy density of the first time-delay eigenstate. A clear focal spot with maximal intensity is seen at the resonator's location.
(e) Upon injecting the second time-delay eigenstate, the strong focusing effect is not seen.
(f) Comparison of the focused amplitude using a phase-conjugation technique (black line), the first time-delay eigenstate (blue line) and the first (red dashed line) and last (black dashed line) reflection eigenchannels.}
\end{figure*}

\subsection*{Coupling to a single resonator}
We begin by demonstrating optimal focusing on a single high-$Q$ dielectric cylinder with a permittivity $\epsilon\sim37$ \cite{Laurent2007}, embedded in a complex environment. As shown in Fig.~\ref{fig1}, the latter is a quasi-two-dimensional multimode waveguide (in the considered frequency range) that is filled with 30 randomly placed low-$Q$ scatterers (teflon cylinders, $\epsilon\sim2.07$). One waveguide end is covered with absorbing foam to mimic open boundary conditions while an array of $N=8$ antennas is located at the other end. As detailed in the Method section and SI, the radiofrequency chain behind each antenna is designed to allow simultaneously the in-situ injection of waves with tailored amplitude and phase profile and the reception of the return signals. These unique capabilities make this microwave setup an ideal candidate for a proof-of-concept demonstration. 

First, we measure the reflection matrix $r(\omega)$ associated with the antenna array between 13 and 14 GHz. 
Second, we apply the WSO $Q_r$ to $r(\omega)$.To compute $r^{-1}(\omega)$, we truncate the last singular value of $r(\omega)$ to prevent experimental noise from corrupting the pseudo-inverse calculation. The spectrum of the real part of $\tilde \tau_n$ is shown in Fig.~\ref{fig1}(b) and a peak is observed on the delay time of the first eigenstate at $f_0=13.58$ GHz. The delay time $\tilde{\tau}_1(f_0)$ reaches 21.5 ns and clearly dominates the other contributions that do not exceed 16 ns. This implies that the resonator's $Q$-factor is $Q_m\sim1850$.

Third, we inject in-situ the normalized eigenvector of the WSO corresponding to the largest delay time at $f_0$. To measure the spatial distribution of the intensity within the medium, we scan the excited field in the scattering medium with a minimally invasive antenna inserted via small holes in the waveguide's top plate (see SI appendix for details). The result shown in Fig.~\ref{fig1}(d) evidences strong focusing at the resonator's location. Relative to the average intensity at that location for the other eigenvectors, the intensity is enhanced by a factor of 10.2. 

We also inject the other time-delay eigenstates (TDEs) into the system. The obtained intensity distribution for the second time-delay eigenstate is shown in Fig.~\ref{fig1}(e), the other field maps are provided in the SI appendix. The intensity at the resonator's location is slightly stronger than for the surrounding background as a consequence of the high $Q$-factor of the resonator, but the incoming wavefront does not result in a proper focal spot. This is confirmed by spectra of the amplitude at the resonator's location for the first seven TDEs in Fig.~\ref{fig1}(c).

In Fig.~\ref{fig1}(f) we benchmark the achieved focusing amplitude with our blind non-invasive scheme against the optimal value attainable with an invasive phase-conjugation scheme. The latter is known to yield the maximum achievable intensity at a selected point by phase-conjugating the field coefficients between the channels and the scanning antenna inserted via the hole above the resonator \cite{Vellekoop2008a}. 
Our proposed scheme achieves $94\ \%$ of the benchmark intensity obtained with an invasive approach. We attribute the slight difference to the non-homogeneous energy density distribution within the resonator.   

We now compare the focused amplitude to the first and last reflection eigenchannels in Fig.~\ref{fig1}(f). In the single scattering regime, the first eigenvector of the matrix $r^\dagger(\omega)r(\omega)$, known as the time-reversal operator \cite{Prada1994}, would also provide focusing on the strongest scatterer in the medium, here, the resonator. However, the correspondence between reflection eigenchannels and scatterers fails in the multiple scattering regime \cite{Aubry2009}. The first eigenchannel mainly excites the first scatterers placed between the antennas and the resonator. 
We also observe that the intensity is not focused on the resonator by exciting the last eigenchannel of $r^\dagger(\omega)r(\omega)$ which corresponds to minimal reflections. For systems with perfectly controlled openings, minimizing the outgoing intensity coherently enhances absorption within the medium \cite{Chong2011} so that scatterers with largest $Q$-factors and hence largest absorption rates may be preferentially excited \cite{Li2018}. Coherent perfect absorption may even be obtained in disordered media \cite{Wan2011,Baranov2017,Pichler2019}. However, we control only a small fraction of incoming and outgoing channels since the system is fully opened at the right side. The eigenchannel with minimal reflection is therefore mainly associated with an increase of transmission from the left to the right so that the intensity on the resonator remains small, as seen in Fig.~\ref{fig1}(f).    

\begin{figure}
\includegraphics [width=8.6cm]{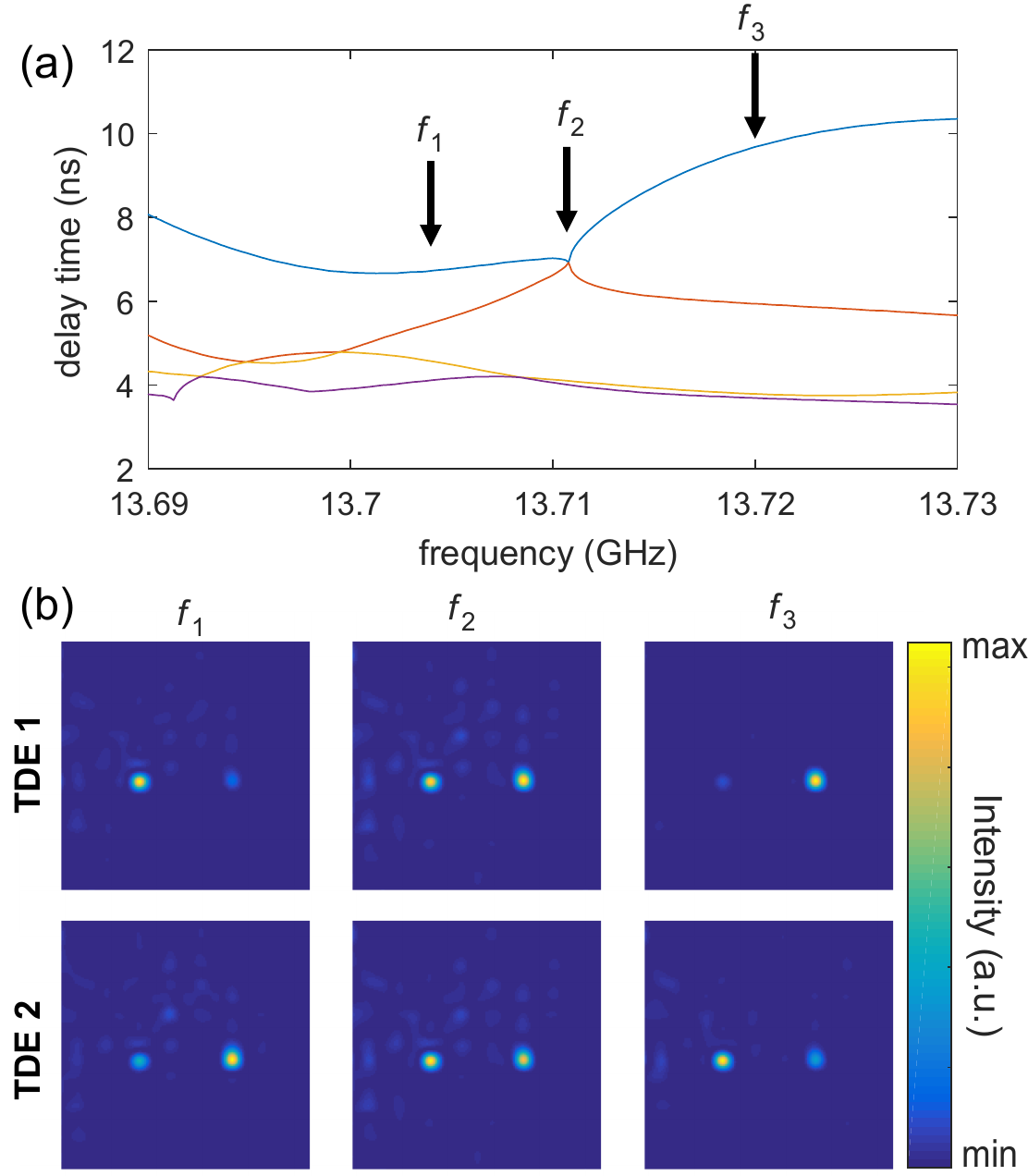}
\caption{Time-delay eigenstates for two embedded dielectric resonators. (a) Spectrum of the real parts of the four largest time-delay eigenvalues. (b) Spatial intensity distributions of the first (top row) and second (bottom row) eigenstates at the three frequencies indicated with arrows in (a).}
\label{fig2}
\end{figure}

\subsection*{Coupling to multiple targets}

Next, we test our approach for a scenario with multiple resonant targets embedded in a scattering environment. Using two identical dielectric cylinders as resonators, we follow the same procedure as before, here between 13.69 and 13.73 GHz. In Fig. \ref{fig2}(a), we observe that the two largest eigenvalues of the WSO clearly stand out, with a crossing at $f_c=13.71$ GHz. By successively injecting the corresponding eigenvectors, we obtain the intensity distributions shown in Fig. \ref{fig2}(b), revealing selective focusing on these two resonators, with a change between positions of maximal intensity at the crossing between the eigenvalues. The eigenstate with largest eigenvalue indeed reveals maximal focusing on the left resonator for $f_0<f_c$ and on the right resonator for $f_0>f_c$. At $f_0=f_c$, the two distributions are the same as a consequence of the hybridization of the two eigenstates.

\subsection*{Coupling to an extended resonator}
\begin{figure}
\includegraphics[width=8.5cm]{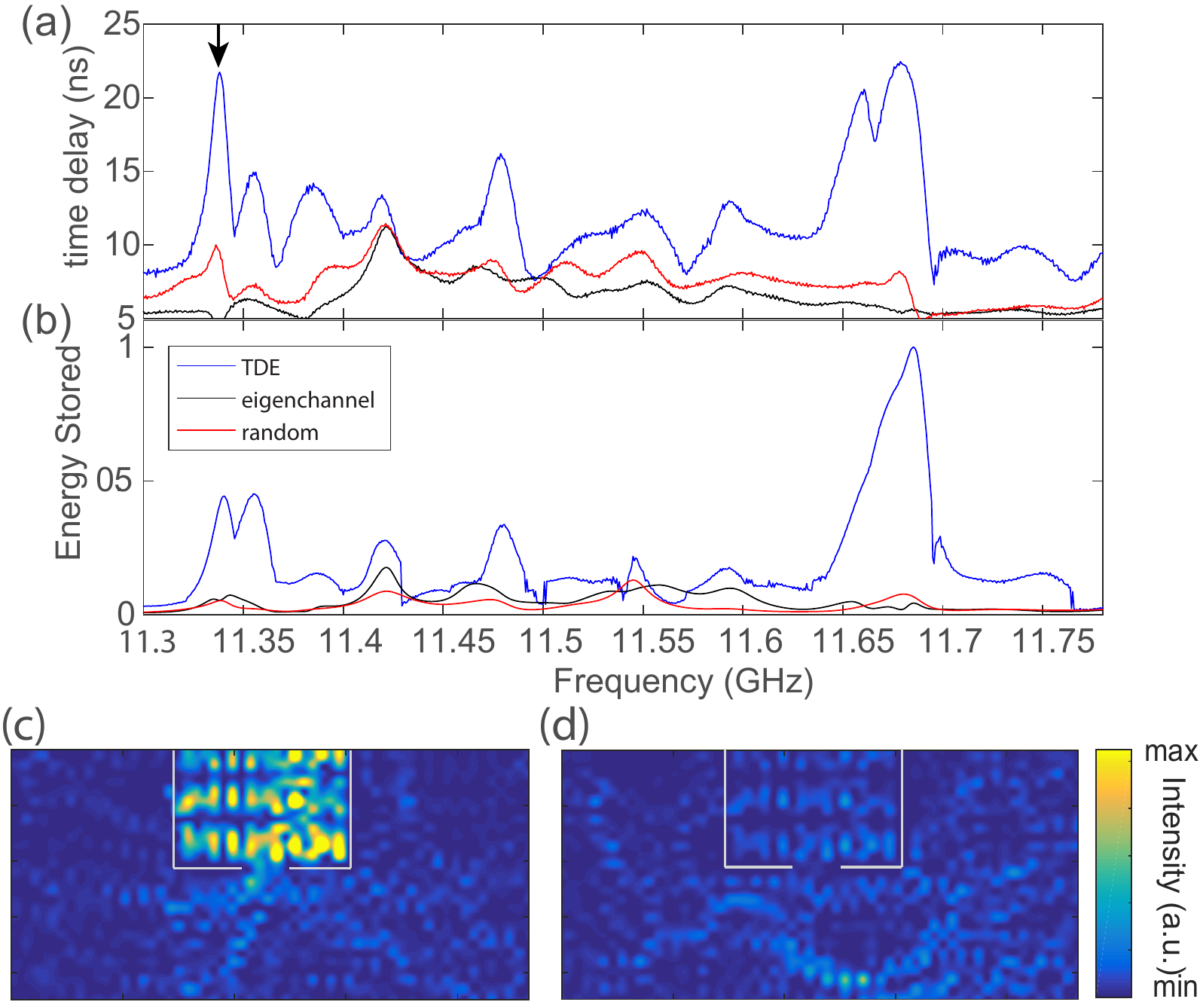}
\caption{\label{fig3} Energy storage in an embedded extended leaky cavity resonator. (a,b) Spectra of the delay time (a) and energy stored within the cavity (b) for the first time-delay eigenstate, the first transmission eigenchannel and a random incoming wavefront. (c,d) Intensity distributions within the medium at $f=11.34$ GHz for the first time-delay eigenstate (c) and the first transmission eigenchannel (d), respectively.}
\end{figure}

We now consider extended resonators with dimensions greater than the diffraction limit. For extended resonators, identifying the wavefront that optimally couples to the resonator is non-trivial even without a surrounding scattering medium. Our extended resonator is a rectangular leaky cavity ($L_c=104$ mm, $W_c=152$ mm) with aluminum walls and an opening of $25 \ \mathrm{mm} \sim 1.05\lambda$ at 11.5 GHz. At the same time, to demonstrate the versatility of our approach, we now work with $t(\omega)$ rather than $r(\omega)$. To that end, we replace the absorbing foam on one end of the waveguide with another array of $N=8$ antennas. We place small pieces of absorbing material in front of the metallic walls between all neighboring antennas to prevent the waveguide from having strong internal reflections (see also discussion below).

We thus compute the eigenstates of the WSO applied to the TM, $Q_t=-it^{-1}(\omega) \frac{\partial t(\omega)}{\partial \omega}$. We obtain the corresponding intensity distributions displayed in Fig. \ref{fig3}, this time by analytically injecting the eigenvectors of the WSO into a second transmission matrix linking the input ports to grid positions within the sample. We calculate the energy stored within the resonator for each eigenstate, $U_n(\omega)$, by integrating the field intensity over the surface of the resonator. For the first time-delay eigenstate, the variation of $U_1(\omega)$ with frequency is seen in Fig. \ref{fig3}(b) to be highly correlated (similarity coefficient 0.68) with the variation of the real part of $\tilde \tau_1$. This correlation highlights the correspondence between the delay time and the energy stored within the cavity.

The enhancement of the stored energy is confirmed by the intensity distribution in the first time-delay eigenstate in Fig. \ref{fig3}(c). The wave is seen to strongly penetrate into the resonant cavity with a spatial distribution of the energy density which is reminiscent of an eigenfunction of a regular cavity. As shown in the theoretical analysis, the incoming wavefront indeed maximally excites the resonant mode at its resonance. The correspondence between the time-delay eigenstates for peaks in $\tilde \tau_1$ and the modes of the cavity at the same frequencies is confirmed in the SI appendix. 

In contrast, upon injecting the first eigenchannel of $t^\dagger (\omega)t(\omega)$, as seen in Fig. \ref{fig3}(d), the wave follows scattering paths around the cavity because an increase of the delay time would also lead to an enhancement of absorption within the cavity. This enhanced absorption would consequently decrease transmission through the sample which is not compatible with a maximization of transmission.

\subsection*{Limitations}

\begin{figure}
\includegraphics[width=8.5cm]{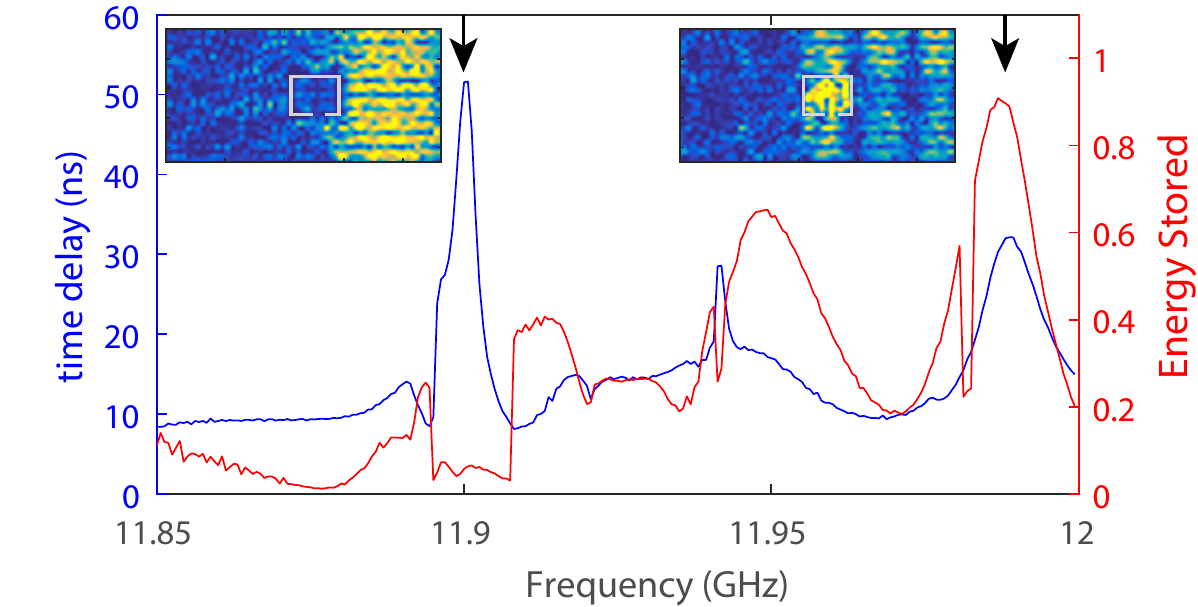}
\caption{\label{fig4} Delay time and energy stored within the cavity for the first time-delay eigenstate. The dimensions of the cavity resonator are now $L_c=56$ mm, $W_c=72$ mm and its aperture is equal to $15$ mm. In the insets, the intensity distributions of the eigenstate at $f_0=11.905$ and $f_0=11.98$ are shown.}
\end{figure}

We finally consider the limitations of our blind-focusing technique. As stated earlier, we assumed up to now that the lifetime of waves in the resonant target is significantly larger than in the surrounding scattering medium. Hence, peaks in the first eigenvalue of the WSO could be identified as a signature of a resonator. If, however, the lifetimes in resonator and medium become comparable, peaks in $\tilde{\tau}_1$ may be associated with eigenstates that are not focusing in the cavity but rather excite modes located outside the cavity. To illustrate this effect, we reduce the size of the cavity placed in the middle of the waveguide (see Fig.~\ref{fig4}) to decrease its coupling to the antennas. Moreover, we remove the pieces of absorbing foam between neighboring antennas at the two waveguide ends so that strong internal reflections appear in the system due to metallic boundary conditions between the openings.

The resulting delay time $\tilde\tau_1$ and energy stored within the cavity resonator $U_1(\omega)$ for the first eigenstate are shown in Fig.~\ref{fig4}. We observe that the peak in $\tilde\tau_1$ at $f_0=11.98$ GHz still corresponds to a  peak of $U_1(\omega)$. However, such a mapping is not observed systematically here. For instance, the first eigenstate at $f_0=11.905$ GHz corresponds to a mode trapped between the top and bottom boundaries of the waveguide which very weakly penetrates into the cavity resonator. The lifetime of this mode largely exceeds the lifetime of waves within the cavity resonator at this frequency. Overall, the degree of correlation between the spectra of $\tilde\tau_1$ and $U_1(\omega)$ is now only 0.33. By closing the waveguide at both ends, the linewidths of resonances in the waveguide have been strongly reduced and the assumption that the medium's linewidths exceed the resonator's linewidth is not valid anymore (see SI appendix for details and simulations). Limitations of our method hence arise for resonators located in high $Q$-factor environments.

\section*{Conclusion}

We have experimentally demonstrated optimal blind-focusing on resonant inclusions in complex scattering environments by controlling delay times of transmitted and reflected waves in a multi-channel system. We reported selective focusing on multiple dielectric resonators as well as on an extended cavity resonator and shed light on the limitations of the scheme when the quality factor of the resonator(s) is not superior to that of the medium. Our approach demonstrated in the microwave range can be extended to optics, acoustics and seismology. We expect these results to trigger new schemes to enhance energy harvesting and non-linear effects in photonic and phononic materials. Our framework may also open new perspectives for deep-imaging through highly scattering samples.

\section*{Methods}
\subsection*{In-situ microwave realization}
A detailed schematic of the radiofrequency chain behind each coax port is provided in Fig.~S1 of the SI appendix. 
A signal, generated by the vector network analyzer's (VNA) transmit port, is equally split into $N=8$ ways with a power divider. Each way is then individually modulated in amplitude and phase by an IQ modulator (IQM). The IQM output is connected to the first port of a 3-port circulator. The modulated signals are injected into the system via the circulator's second port. 
Simultaneously, the return signal from the system enters and exits the circulator via its second and third port, respectively. The $N$ return signals for a given incoming wavefront are measured by connecting the third port of each circulator to an $N\times1$ electromechanical switch that is in turn connected to the VNA's receive port.
To measure the reflection matrix, we select one incoming port at a time by setting the modulation of its IQM to unity and the remaining ones to zero.

The field within the scattering medium is scanned non-invasively using a wire antenna which is inserted into a grid of holes (diameter: 4 mm, spacing: 8 mm) that are drilled into the waveguide’s 6-mm-thick top plate. The wire length is chosen to coincide with the top plate's thickness so that the wire does not penetrate into the waveguide.

\subsection*{Computation of time-delay eigenstates}
Applying the WSO to an incomplete scattering matrix, for instance, the RM or TM, complicates the evaluation of the necessary matrix inversion. Taking the example of $Q_t(\omega)$, we first decompose $t$ into singular values, $t = U \Sigma V^\dagger$, and then use its Moore-Penrose pseudoinverse, $t^{-1} = V \Sigma^{-1} U^\dagger$. $\Sigma$ is the diagonal matrix of singular values and $U$ and $V$ are the unitary matrices of left and right singular vectors of $t$. We observe that including the last singular value makes the WSO unstable due to experimental noise and the non-vanishing frequency shift used to compute the derivative of $t(\omega)$ with respect to $\omega$. We therefore apply a singular-value truncation on $t$ before estimating $t^{-1}$. Specifically, we do not take into account the last singular value, being the one that is most easily corrupted in experiments.

\section{Author contributions}
The project was initiated and conceptualized by M.D. The experiments were carried out by P.d.H and M.D. and the simulations were performed by R.S. and M.D. All authors thoroughly discussed the results. The manuscript was written by P.d.H. and M.D. and reviewed by all authors.

\section*{Acknowledgements}

This publication was supported by the European Union through the European Regional Development Fund (ERDF), by the French region of Brittany and Rennes Métropole through the CPER Project SOPHIE/STIC \& Ondes and by the French “Agence Nationale de la Recherche” under reference ANR-17-ASTR-0017. We would also like to acknowledge C. Leconte for her help in automating the scan and P.E. Davy for the 3D rendering of the experimental setup.

\bibliography{pnas-sample}

\newpage
\beginsupplement

\section{Supplementary Information}

\section{Technical details of the microwave in-situ experiments}
\begin{figure}[!htb]
\centering
\includegraphics [width=8.5cm]{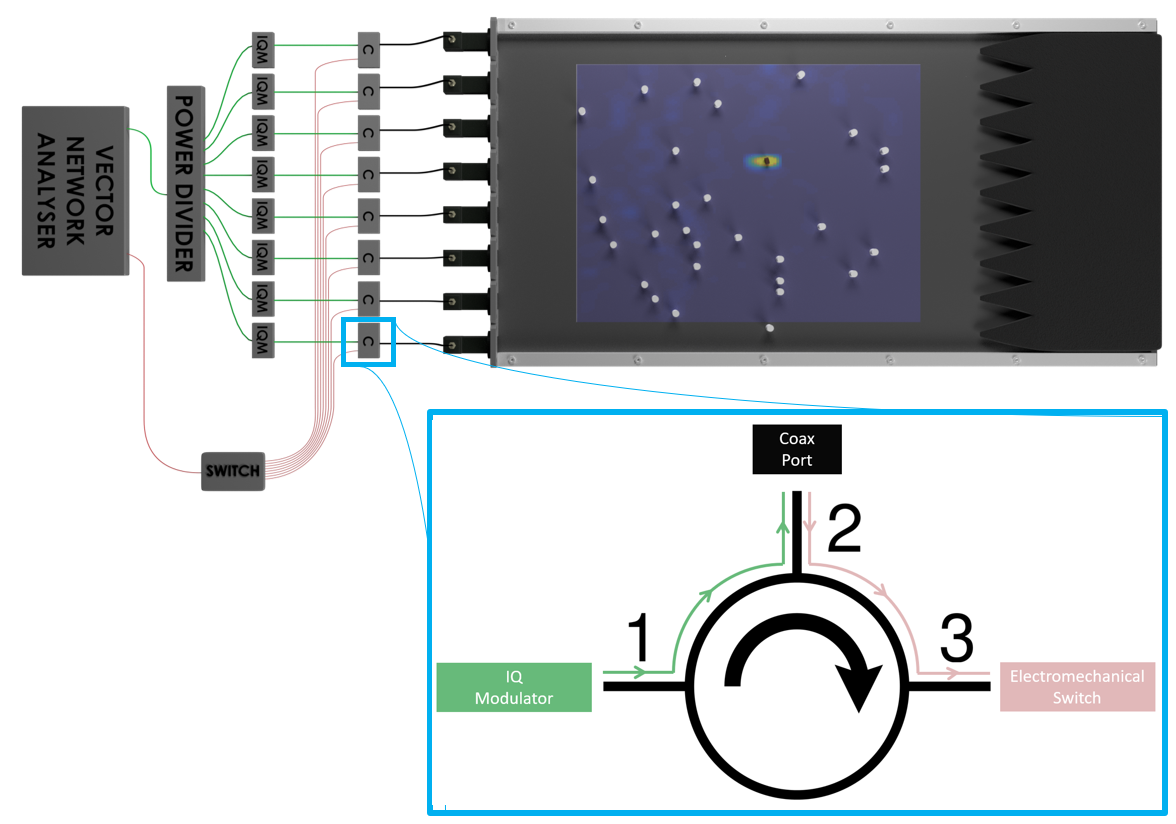}
\caption{Detailed schematic drawing of the experimental setup. The radiofrequency chain behind each of the eight coax antennas is designed to allow simultaneously the in-situ injection of waves with tailored amplitude and phase profile and the measurement of the return signal. The inset clarifies the role of the circulators in this setup. See Materials and Methods in the main text for details.}
\label{experimental_setup}
\end{figure}

Fig.~\ref{experimental_setup} displays the technical details of the microwave in-situ realization described in the Methods section of the main text.

\section{Detailed theoretical proof of optimality}

In this section, we analytically apply the Wigner-Smith time-delay operator (WSO) to a measurement matrix for a resonator surrounded by a scattering environment. For concreteness, we consider the case of the reflection matrix (RM) $r(\omega)$ but the following calculations can be generalized to other measurement matrices such as the transmission matrix (TM) or the complete scattering matrix. 

As motivated in the main text, we decompose the RM $r(\omega)$ as

\begin{equation}
r(\omega)=r_0(\omega)+r_n(\omega),
\end{equation}

\noindent where $r_0(\omega)$ corresponds to reflected waves that did not interact with the resonator and $r_n(\omega)$ is the contribution of waves that did interact with the resonator.
Note that $r_0(\omega)$ and $r_n(\omega)$ include multiple scattering effects within the medium. The frequency $\omega$ is chosen to be close to the central frequency $\omega_n$ of a long-lived resonance of the resonator. The linewidth of the resonance is denoted $\Gamma_n$. The contribution of this mode to the RM is a matrix of rank one given by \cite{Rotter2001,Davy2019}

\begin{equation}
r_n(\omega)=-i\frac{W_n W_n^T}{\omega-\tilde\omega_n}.
\end{equation}

\noindent This equation involves the complex eigenfrequency $\tilde\omega_n=\omega_n-i\Gamma_n/2$ and the vector $W_n$ of dimension $N\times1$ which is the projection of the eigenfunction onto the channels of the RM. $W_n$ is independent of frequency for a high-$Q$ resonance with narrow linewidth. 

We now calculate the inverse of $r(\omega)$. Because $r_n(\omega)$ is a matrix of rank one, we can apply the Sherman–Morrison formula \cite{Sherman1950} which yields 

\begin{equation}
r^{-1}(\omega)=r_0^{-1}(\omega)-\frac{r_0^{-1}(\omega)r_n(\omega) r_0^{-1}(\omega)}{1+\kappa(\omega)}.
\end{equation}

\noindent The complex-valued scalar $\kappa(\omega)$ is equal to the trace of the matrix $r_0^{-1}(\omega)r_n(\omega)$:

\begin{equation}
\kappa(\omega)=\mathrm{Tr}(r_0^{-1}(\omega) r_n(\omega))
= -i\frac{ W_n^T r_0^{-1}(\omega) W_n }{\omega-\tilde\omega_n}.
\end{equation}

\noindent Assuming that the delay time of the reflected waves is dominated by the waves interacting with the resonator implies that the variation of $r_0(\omega)$ with frequency is much smaller than the variation of $r_n(\omega)$ with frequency, i.e. $\frac{\partial r_0(\omega)}{\partial \omega} \ll \frac{\partial r_n(\omega)}{\partial \omega}$ so that $r_0(\omega)=r_0$. This yields

\begin{equation} 
\frac{\partial r(\omega)}{\partial \omega} \approx \frac{\partial r_n(\omega)}{\partial \omega} = i\frac{W_n W_n^T}{(\omega-\tilde\omega_n)^2} = -\frac{r_n(\omega)}{\omega-\tilde\omega_n}\\
\end{equation}

\noindent The WSO $Q_r(\omega)= -i r^{-1}(\omega) \frac{\partial r(\omega)}{\partial \omega}$ can hence be evaluated as

\begin{equation} 
Q_r(\omega)=-ir^{-1}(\omega)\frac{r_0-r(\omega)}{\omega-\tilde\omega_n}=\frac{-i}{\omega-\tilde\omega_n}(r^{-1}(\omega)r_0-\mathbb{1}),\\
\end{equation}

\noindent where $\mathbb{1}$ is the identity matrix. Using that 

\begin{equation} 
r^{-1}(\omega)r_0=\left(r_0^{-1}-\frac{r_0^{-1}r_n r_0^{-1}}{1+\kappa(\omega)}\right)r_0=\mathbb{1}-\frac{r_0^{-1}r_n}{1+\kappa(\omega)},\\ 
\end{equation}

\noindent this leads to

\begin{equation} 
Q_r(\omega)=i \frac{1}{\omega-\tilde\omega_n} \frac{r_0^{-1}r_n(\omega)}{1+\kappa(\omega)}.\\
\end{equation}

\noindent By inserting the expression of $r_n(\omega)$, this can equivalently be written as 
\begin{equation} 
Q_r(\omega)=\frac{1}{(\omega-\tilde\omega_n)^2} \frac{r_0^{-1}W_n W_n^T}{1+\kappa(\omega)}.
\end{equation}

\noindent This expression provides a framework for the analysis of the WSO in the case of non-resonant contributions to the measured matrix. We observe that the matrix $r_0^{-1}W_n W_n^T$, and hence $Q_r(\omega)$, is again of rank one since $r_0^{-1}W_n$ is a $N\times1$ vector. This implies that the left eigenvector of $Q_r(\omega)$, i.e. the vector of dimension $N\times1$ verifying $q_n^T Q_r=\tilde\tau_n q_n^T$, is $q_n=W_n/||W_n||$. Its conjugate $q_n^*$ gives the wavefront which is the time-reversed of the coupling wavefront between the mode and the channels. This wavefront maximizes the contribution of the mode and is therefore the wavefront for maximal coupling to the resonator's mode \cite{Davy2019}. In other words, it is the optimal wavefront for focusing on the resonator so that the energy stored within the resonator will be maximized.

The associated eigenvalue $\tilde\tau_n$ is equal to the trace of the WSO 
\begin{equation} 
\tilde\tau_n=\frac{i}{\omega-\tilde\omega_n} \frac{\kappa(\omega)}{1+\kappa(\omega)}\\.
\end{equation} 

\noindent The expression can be simplified at the resonance with the quasi-normal mode, $\omega=\omega_n$, 

\begin{equation} 
\tilde\tau_n=\frac{2}{\Gamma_n} \frac{\kappa(\omega_n)}{1+\kappa(\omega_n)}\\
\end{equation} 

\noindent This eigenvalue hence mostly depends on the coupling strength between the non-resonant and resonant part through the complex parameter $\kappa(\omega_n)=-2W_n^Tr_0^{-1}W_n/\Gamma_n$.

In the limit $|\kappa| \gg 1$, the eigenvalue is real and is equal to $\tilde\tau_n=2/\Gamma_n$. The eigenvalue therefore directly provides the delay time associated to the quasi-normal mode. $|\kappa| \gg 1$ corresponds to the case of a small contribution of the non-resonant term $r_0$. In the limit $|\kappa| \ll 1$, the first eigenvector still gives the wavefront for optimal coupling to the resonator but the associated eigenvalue is not anymore a well-defined delay time.

\section{Spatial intensity distributions of TDEs for coupling to a single resonator}

\begin{figure} [!htb]
\centering
\includegraphics [width=8.5cm]{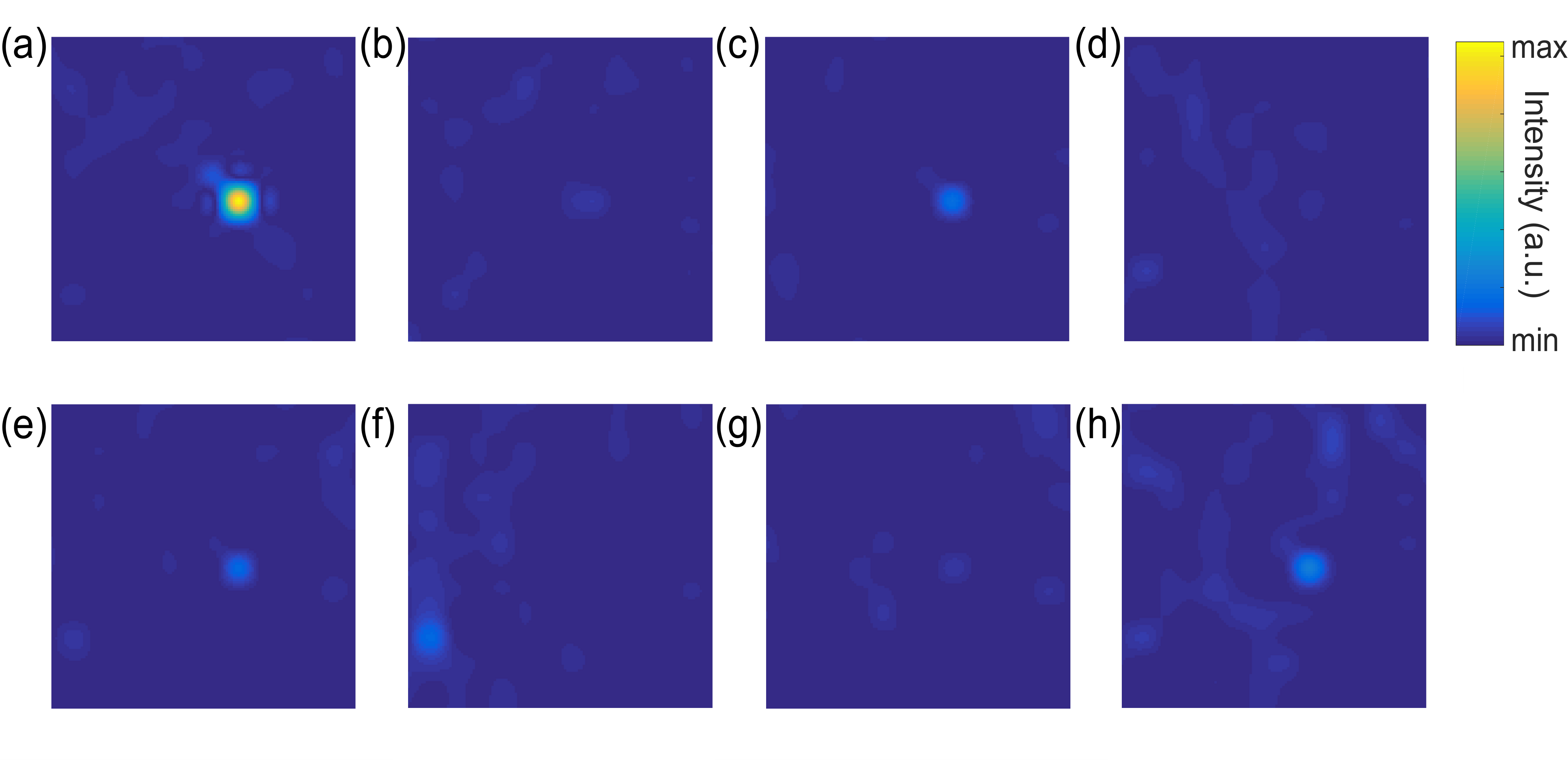}
\caption{Spatial intensity distribution for the eight time-delay eigenstates corresponding to the case of a single dielectric resonator embedded in the scattering medium. Note that (a,b) are identical to Fig.~1(d,e) in the main text.}
\label{8TDE}
\end{figure}

In Fig.~\ref{8TDE} we present the eight spatial intensity distributions corresponding to the eight time-delay eigenstates. Note that Fig.~\ref{8TDE}(a,b) are identical to Fig.~1(d,e) in the main text.
The intensities are normalized by the maximum in Fig.~\ref{8TDE}(a). Focusing on the resonator is clearly observed in Fig.~\ref{8TDE}(a) for the first TDE. We note that the intensity is still slightly enhanced on the maps of the other TDEs, even though the wave is not focused. This is a consequence of the resonator's high $Q$-factor which leads to an enhanced energy density relative to the background even for a random incident wavefront. 

\section{Simulations}

In this section, we show in finite-element simulations the correspondence between quasi-normal modes with small linewidths and the first time-delay eigenstate. 

\subsection{Simulation setup} 
\begin{figure} [!htb]
\centering
\includegraphics [width=8.5cm]{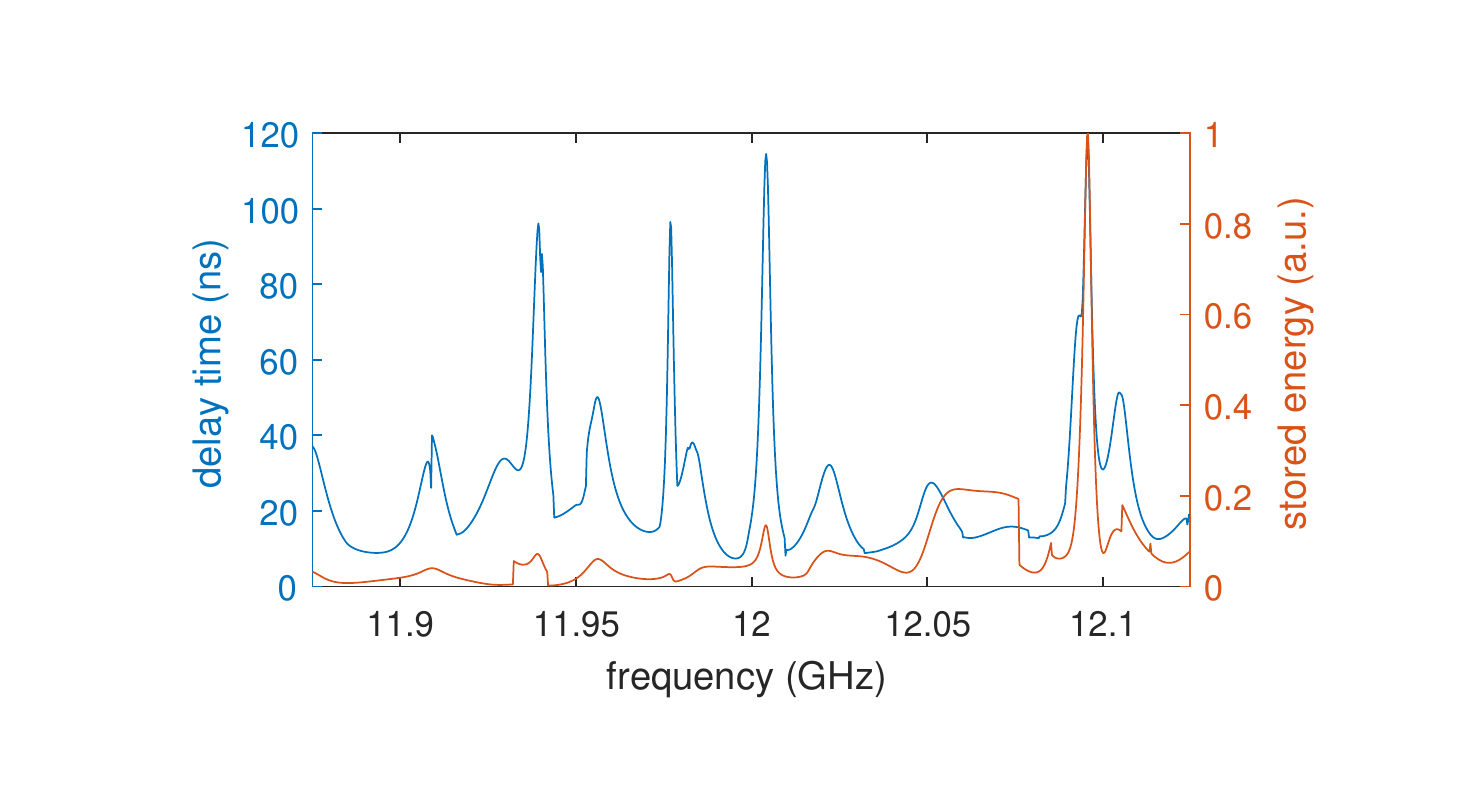}
\caption{(a) Sketch of the two-dimensional setup in simulations. A leaky cavity and 20 teflon cylinders are placed within the system. (b) Spectra of the first proper delay time (first eigenvalue of the WSO, blue line) and the energy stored within the resonator for the first time delay eigenstate (orange line).} 
\label{spectrum}
\end{figure}

The two-dimensional geometry of the experimental setup was reproduced with the COMSOL software -- see Fig.~\ref{spectrum}(a). The aluminum boundaries in the experimental setup are replaced with perfect electric conductors. This includes the spacing between antennas at the left and right sides. We do not include homogeneous losses here so that the lifetime of waves within the sample is much stronger in simulations than in experiments. The antennas are modelled by single-mode ports. 20 teflon cylinders of radius equal to 3 mm are randomly placed inside the waveguide.

The resonator is a leaky rectangular cavity: a square with dimension 7$\lambda_0$/2. The aperture of this cavity is equal to $\lambda_0$. Simulations are performed between 11.875 and 12.125 GHz on 1000 frequency points, corresponding to a frequency step $\Delta_f=250$ kHz and a wavelength $\lambda_0$ = 25 mm.

\subsection{Using the full scattering matrix}

We extract spectra of the $16\times16$ scattering matrix and find the proper delay times which are the eigenvalues of the WSO. Because the openings of the cavity are fully controlled, the scattering matrix is complete and hence unitary. As a result, the eigenvalues $\tau_n(\omega)$ of $Q(\omega)$, known as the proper delay times, are real. 
 Fig.~\ref{spectrum}(b) displays the spectrum of the first proper delay time $\tau_1(\omega)$ which is the first eigenvalue of the WSO. We observe that $\tau_1(\omega)$ presents multiple peaks. We identify these peaks as the correspondence between the first time-delay eigenstate and the resonances of the medium. The field corresponding to the time-delay eigenstate is found using the linearity of the wave equation by coherently summing the spatial distribution of the fields within the medium for each transmitting antenna. The eigenfunctions of the system are also obtained in COMSOL simulations using the eigenvalue solver. 57 eigenfrequencies are found in the considered frequency range.

\begin{figure}[!htb]
\centering
\includegraphics [width=8.5cm]{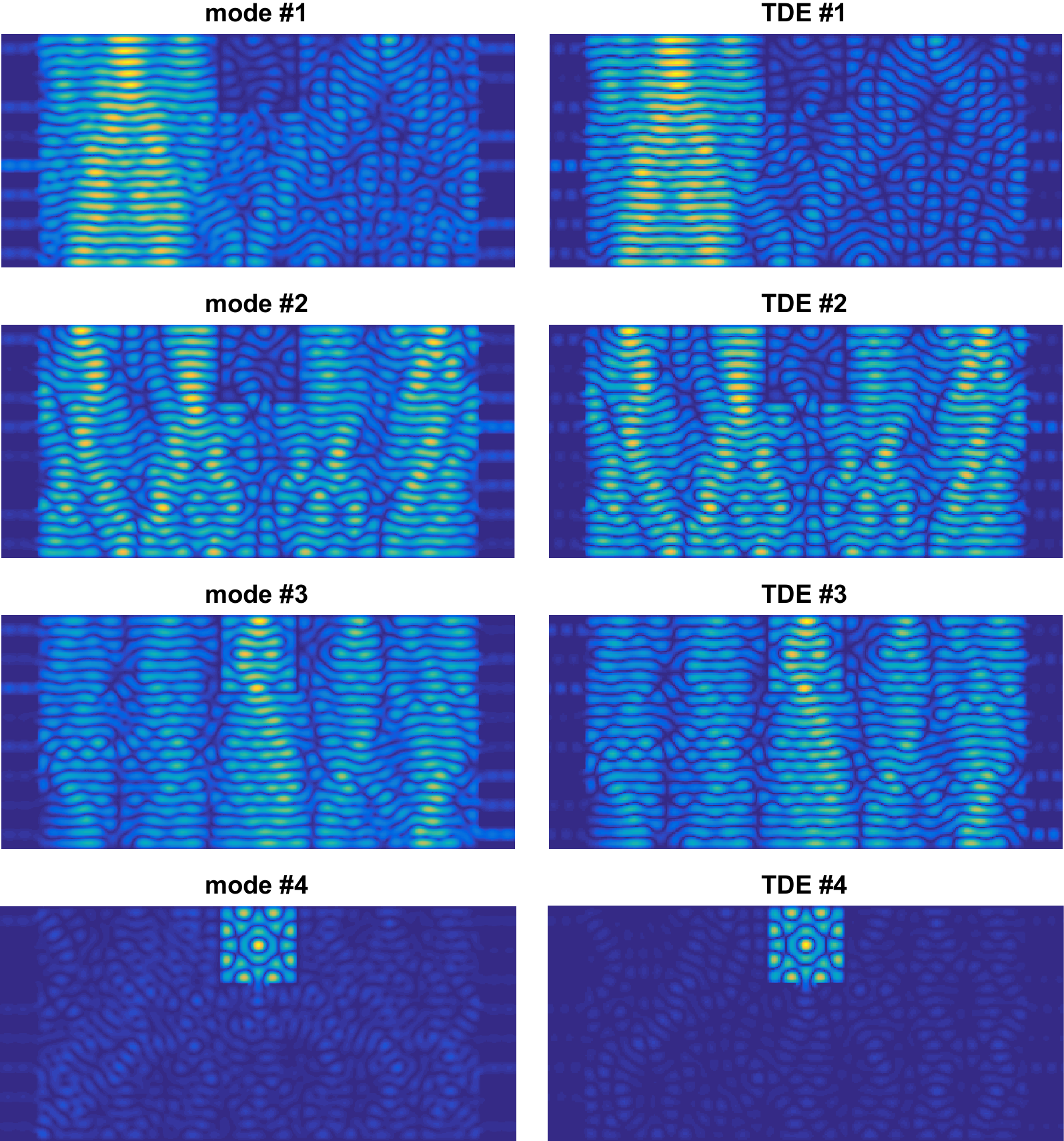}
\caption{(Left column) Eigenfunctions of quasi-normal modes with central frequencies at $f_1=11.94$ GHz, $f_2=11.98$ GHz, $f_3=12.01$ GHz and $f_4=12.09$ GHz. (Right column) Spatial distribution of time-delay eigenstates found at $f_1$, $f_2$, $f_3$ and $f_4$.} 
\label{modes}
\end{figure}

For the four peaks corresponding to the four arrows in Fig.~\ref{spectrum}(b), the eigenfunction of the quasi-normal mode which is at resonance is very close to the corresponding time-delay eigenstate. This correspondence is explained in the main text and in Section 2 of this supplementary information using the decomposition of the WSO as a superposition of modal contributions. At resonance with a spectrally peaked resonance, $\tau_1(\omega)\sim 2/\Gamma_n$, where $\Gamma_n$ is the linewidth associated with the resonance.
Given the lack of homogeneous losses in these simulations, the modes with the shortest linewidths are not exclusively associated with the resonating cavity; some also correspond to trapped modes within the system that only weakly excite the cavity. Strong internal boundary condition indeed results from the metallic spacing between the ports, so that some modes can be spatially localized outside the resonator but with small linewidths. This is the case of the three first peaks for which the eigenstates are shown.

However, for modes that are localized within the resonator, the first time-delay eigenstate strongly enhances the energy stored within the cavity, $U_1(\omega)$. The peak on the time delay spectrum at 12.09 GHz (forth arrow in Fig.~\ref{spectrum}(b)) is accompanied by a strong enhancement of the stored energy which is evaluated by integrating the intensity on the surface $\Omega$ of the resonator

\begin{equation} 
U_1(\omega)=\int_\Omega \mathrm{d}x \mathrm{d}y \ |E^T(x,y,\omega)q_1|^2 .
\label{energy_stored}
\end{equation}

\noindent Here $E^T(x,y,\omega)$ is the vector of field transmission coefficients between the incoming channels and the domain $\Omega$ of the resonator, and $q_1$ is the first left eigenvector of $Q(\omega)$.

\subsection{Optimal coupling}

We now benchmark the energy storage achieved upon injecting the first WSO eigenvector as a wavefront against the use of an iteratively optimized wavefront. Using the knowledge of the field within the resonator for each transmitting antenna, we extract in an optimization procedure the incoming wavefront that maximizes $U_1(\omega)$ using a non-linear programming solver. Note that the latter obviously relies on invasive field measurements inside the system. This benchmarking complements the theoretical argument in Section 2 to prove the optimality of our scheme, as well as the comparison in the main text of the focused intensity on a small resonator with the intensity focused using a phase conjugation technique.

\begin{figure} [!htb]
\centering
\includegraphics [width=8.5cm]{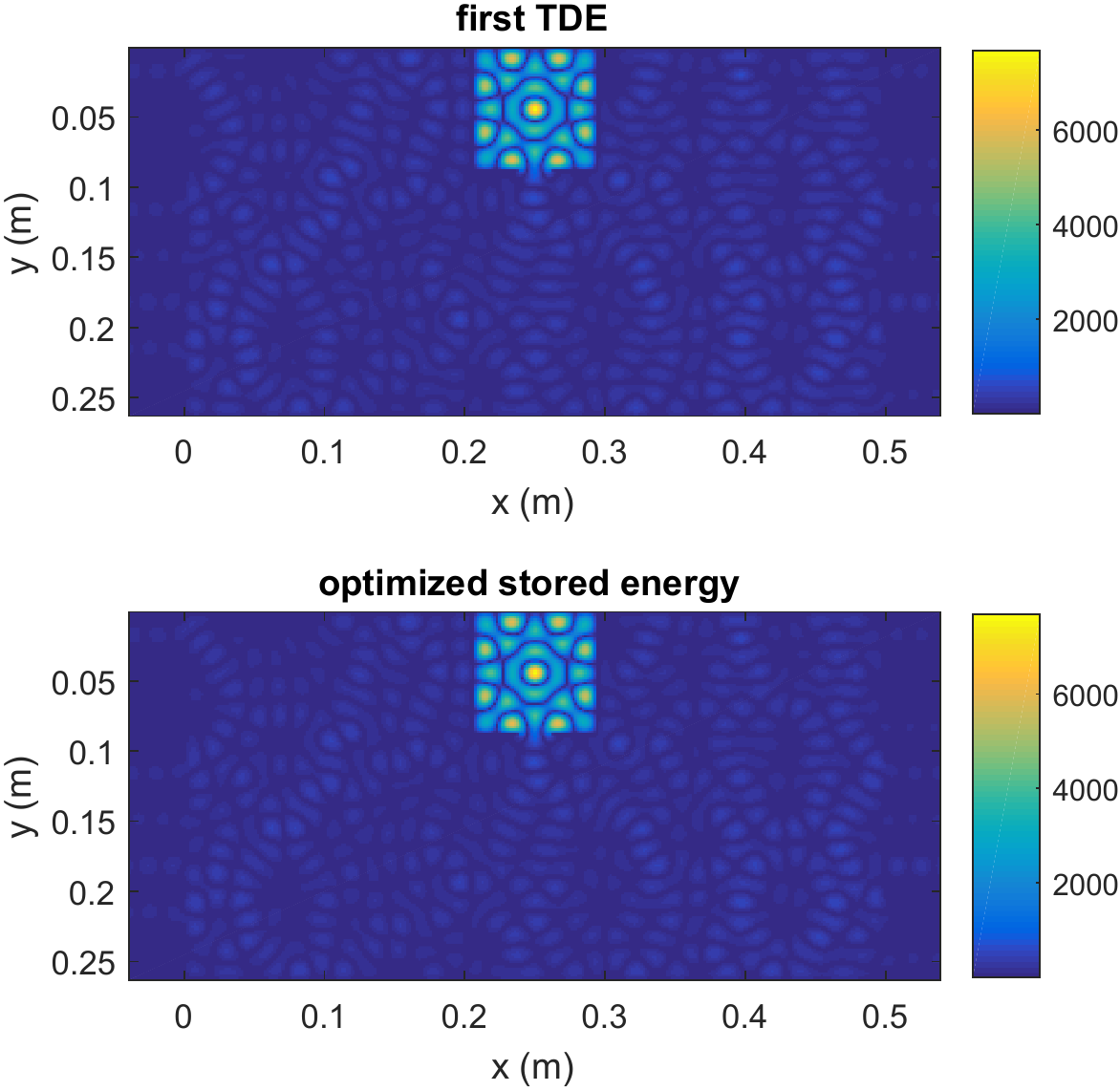}
\caption{(Left column) Spatial distribution of the energy density for (a) the first time delay eigenstate at $f_0$ and (b) an incoming wavefront which maximizes the energy stored within the cavity.} 
\label{optimal_coupling}
\end{figure}

 The two spatial distributions are shown in Fig.~\ref{optimal_coupling} and are seen to be almost identical. The ratio of energy stored between the time-delay eigenstate and its maximized value is 99.5\%. The small difference between the two most probably arises from the non-zero frequency step used to estimate the derivative of the scattering matrix involved in the WSO.  

\subsection{Using only the Transmission Matrix}
\begin{figure} [!htb]
\centering
\includegraphics [width=8.5cm]{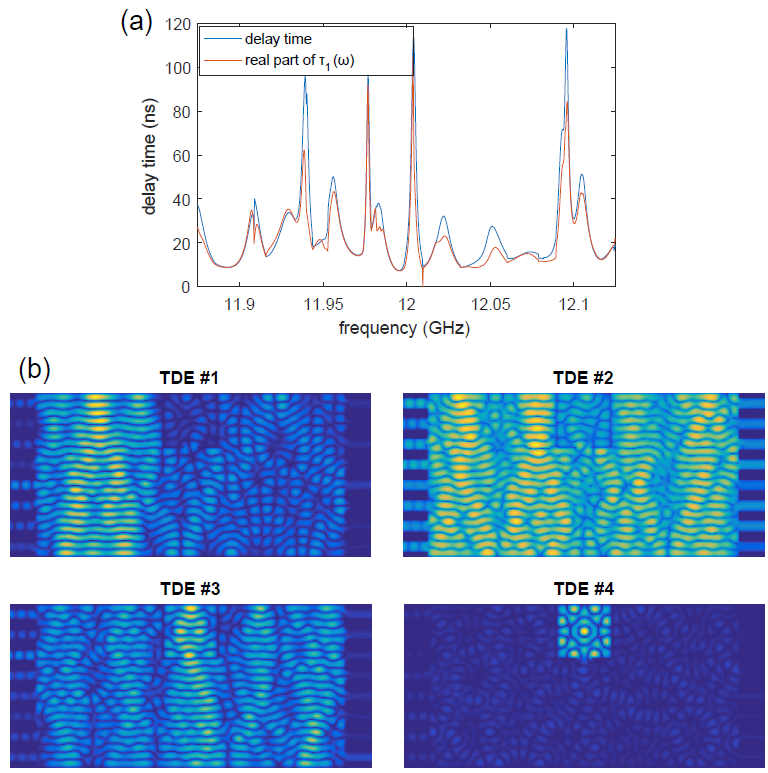}
\caption{(a) Spectra of the delay times associated to first eigenstate of $Q_t$ (blue line) and real part of the first eigenvalue of $Q_t$ (orange line). (b) Spatial distribution of time-delay eigenstates found from the Wigner-Smith operator $Q_t$ at $f_1$, $f_2$, $f_3$ and $f_4$.} 
\label{TM_fields}
\end{figure}

In the following, the WSO is calculated using only the transmission matrix, $Q_t$, as in microwave measurements presented in the second half of the main text. We compare in Fig.~\ref{TM_fields}(a) the real part of the largest time-delay eigenvalue with the delay time found using Eq. (1) of main text with the incoming wavefront being $E_i=q_1$. The two curves follow the same trend, demonstrating that the real part of $\tilde \tau_1(\omega)$ gives a good estimation of the real delay time of waves injected within the system. 

We can again identify the time-delay eigenstates associated to peaks in $\tilde \tau_1(\omega)$ as a signature of quasi-normal modes of the medium. For the fourth peak found at $f_4=12.09$ GHz, the energy density is smaller than upon using the full scattering matrix. Indeed, having exploited only eight channels out of sixteen to focus the energy density, the ratio of stored energy using $t$ and $S$ is found to be $\sim0.7$.

The energy stored in the first time-delay eigenstate is now 95\% of the maximum stored energy found in optimization. However, to estimate $t^{-1}$, the contribution of the last singular value was discarded to avoid unstable results. Only seven subspaces hence contribute to estimate the WSO. This hence obviously reduces the stored energy in comparison to its maximal value found using an invasive technique.


\end{document}